# Shared memories driven by the intrinsic memorability of items

Wilma A. Bainbridge, PhD
University of Chicago

**Abstract**  When we experience an event, it feels like our previous experiences, our interpretations of that event (e.g., aesthetics, emotions), and our current state will determine how we will remember it. However, recent work has revealed a strong sway of the visual world itself in influencing what we remember and forget. Certain items--including certain faces, words, images, and movements-- are intrinsically memorable or forgettable across observers, regardless of individual differences. Further, neuroimaging research has revealed that the brain is sensitive to memorability both rapidly and automatically during late perception. These strong consistencies in memory across people may reflect the broad organizational principles of our sensory environment, and may reveal how the brain prioritizes information before encoding items into memory. In this chapter, I will discuss our current state-of-the-art understanding of memorability for visual information, and what these findings imply about how we perceive and remember visual events.

## 1. Introduction

At all waking moments, we are experiencing a continuous, never-ending flow of sensory information. Even during a normal morning routine, you may be watching the news on TV, scrolling through social media, eating breakfast, and conversing with your family, all simultaneously. While our memories are rich and detailed (Brady et al., 2008; Bainbridge et al., 2019), we usually cannot remember everything (Cowan, 2010). When asked about that morning the following week, some details might be preserved in your memory (e.g., what you wore that day), while others may be completely gone from memory (e.g., what you ate for breakfast). While various cognitive processes will influence what you remember from that day—your emotional state, what you are actively paying attention to, your level of

fatigue—the events themselves will also have a large influence over what you remember. For example, a particularly distinctive news headline may be captured in your memory, even if you are groggy or distracted. The intrinsic power of the stimulus to influence our memories—the intrinsic *memorability* of an event—has become a hot topic in the fields of human perception and memory. A growing body of work aims to understand what factors drive the memorability of certain items over others, and what consistencies in memory across people may imply about the underlying mechanisms of perception and memory.

This chapter will discuss our current understanding of stimulus memorability as it relates to human perception. It will begin in Section 2 with an overview of the concept of memorability, and a how-to guide on how memorability can be quantified behaviorally for any type of stimulus. Next, Section 3 will discuss our current psychological understanding of memorability and how it relates to other perceptual and semantic properties of a stimulus. Finally, Section 4 will present the latest neuroscientific understanding of memorability, with results suggesting that memorability reflects a *prioritization signal* of a perceptual input. Throughout, we will discuss what these findings may imply for the computational exploration of memorability.

## 2. Memorability for visual events

While many would agree that intuitively some images are more memorable to us than others, it is not a given that memorability would be a quantifiable attribute for a given image. It has long been known that memories are highly malleable, idiosyncratic, and dependent on our own personal experiences. Distinct cognitive processes and behaviors have been identified for viewing familiar (i.e., previously experienced) faces, like those of celebrities, versus completely novel faces (Rossion et al., 2003; Eger et al., 2005). For example, our ability to recognize a face from a different viewpoint improves when we are highly familiar with that face (Klatzky and Forrest, 1984; Megreya and Burton, 2006; Jenkins et al., 2011). Relatively low consistency has been observed across observers for many other attributes of an image; for example, participants do not agree on which faces are the most typical, interesting, kind, or even which they think will be most memorable (Bainbridge, 2017). Thus, initially one might expect that memory performance for a given image is almost entirely observer-dependent, and that the stimulus itself may contribute relatively low predictive power.

However, the first studies on memorability revealed a remarkable consistency in the images that people remembered and forgot (Isola et al., 2011; Bainbridge et al., 2013). In spite of our unique individual experiences, in a large and diverse sample of participants tested across the United States, there were some face and scene images that most people remembered, and some that most people forgot (see Figure 1). Importantly, the image itself as a factor contributed to more than half of the variance in memory performance (Bainbridge et al., 2013), implying that the images we view are just as important as our own state and prior experiences in determining what we ultimately remember and forget. Further, these results suggested that we can conceptualize *memorability* as an intrinsic, quantifiable property of an image; an image can be 80% memorable or 20% memorable, and one can use a "memorability score" to predict memory performance for a new set of observers.

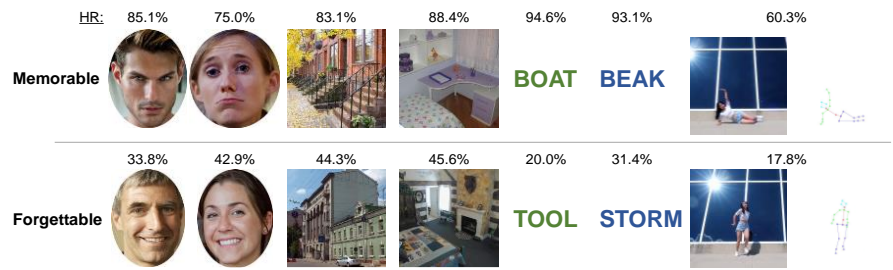

**Fig. 1. Example memorable and forgettable images**. Shown here are example stimuli at the opposite ends of memorability (indicated here as hit rate), in the domains of faces (Bainbridge et al., 2013), scenes (Isola et al., 2011), words (Xie et al., 2020), and dance moves even with visual information about the dancer removed (Ongchoco, submitted).

Since these first findings of consistent memory performance in face and scene images, memorability consistency has been observed across a diverse range of stimulus types (Figure 1). Consistent memorability has been observed in highly dynamic, rich visual stimuli such as videos (Cohendet et al., 2019; Newman et al., 2020), movie scenes (Cohendet et al., 2018), and faces across transformations of viewpoint and expression (Bainbridge, 2017). Consistent memorability has also been observed in highly human-constructed types of visual content, such as visualizations and infographics (Borkin et al., 2013). Conversely, memorability has been observed for stimulus types with relatively low visual information such as words (Xie et al., 2020) and actions across visual formats (Ongchoco et al., submitted). While the current chapter will largely focus on the memorability of images, current evidence suggests that we would find similar patterns across other types of stimuli.

## 2.1 How do we capture and operationalize memorability for a stimulus?

It is relatively straightforward to gather the "ground truth" memorability scores for a set of stimuli. Memorability has been most commonly tested (and quantified) using a *continuous recognition* task, where participants view a stream of images (or videos), and press a button whenever they spot a repeat from earlier in the sequence (Isola et al., 2011). This task is used most commonly because it is relatively time efficient and engaging for the participant—participants are making judgments as each image is presented, and so we can quantify engagement and memory performance in real-time. However, the specific task used is flexible, as memorability scores from a continuous recognition task have also been shown to replicate in paradigms using separate study and test phases (Goetschalckx et al., 2018) and in perceptual tasks that surprise participants with a memory test at the end (e.g., incidental memory paradigms: Bainbridge et al., 2017; Goetschalckx et al., 2019; Bainbridge, 2020).

In conducting a continuous recognition task, there are many important considerations when designing the experiment to collect memorability scores (Figure 2):

1. **The parameters should be determined for the memory test.**

    Memorability effects have been shown to be robust across very different experimental parameters. Even if an image is shown for only 13ms (Broers et al., 2018), or up to 10s (Borkin et al., 2015), its memorability is likely to influence human memory performance. Similarly, gaps between image repeats for as little as 36s (Isola et al., 2013), or as long as a week (Goetschalckx et al., 2017) still result in the same items emerging as memorable or forgettable. One important consideration is that with a continuous recognition task, target images (those for which you are collecting memorability scores) must be separated by filler images in between target repeats. These filler images often consist of "vigilance repeats", repetitions of filler images at the range of 1-5 images apart, to serve as exclusion criteria for participants who are inattentive and not noticing image repeats even at short delays. Generally, a ratio of ¼ targets to ¾ filler is most common, and filler images should not be distinguishable from target images (in terms of visual, semantic, or categorical properties), to prevent participants from acting differently for targets versus fillers. If memorability scores are desired for all items, then the target images should counterbalanced across participants, so that each item serves as a target for a large number of participants.

    Another consideration is how to incentivize high performance in participants. Past studies have incentivized participants to provide as much data as possible, by paying participants based on number of images viewed in the stream (e.g., Isola et al., 2011). Other studies have paid participants a consistent payment amount and only refused payments to participants who did not respond on the task (e.g., Bainbridge et al., 2013). Some studies require participants to perform well on the vigilance task in order to continue with the task (including a requirement to maintain low false alarms; Isola et al., 2011), while others have not required performance minimums (e.g., Bainbridge, 2020). One important note to keep in mind is that the tasks can be rather difficult (e.g., Bainbridge, 2017), and those with poor memory should not necessarily be punished. Introducing a reward for high memory performance could also alter memory performance, although thus far reward has shown no interaction with stimulus memorability on memory performance (Bainbridge, 2020). Thus far, data quality and across-participant consistency has been relatively high, even

without the introduction of high rewards or punishments (e.g., Bainbridge, 2020).

For researchers wanting to create their own memorability experiments, I have created a publicly available online tool[1] that can generate the entirety of the code for a memorability experiment, given a few inputs from the researcher on these parameters of the study.

## 2. The experiment should be run with a large and diverse sample of participants online.

Memorability experiments have been largely run online (on platforms such as Amazon Mechanical Turk[2], AMT), because of the rapid access to large numbers of diverse participants. Running these experiments in a smaller convenience sample (e.g., with local college students) may hamper the generalizability of the results; it would be impossible to know whether consistencies in memory performance exist regardless of diverse and broad experiences across people, or due to commonalities in that narrow participant sample (e.g., a face that looks like a university's Dean may be highly memorable to that university's students, but not to people outside that university). It is also important to collect a large number of memory responses per stimulus, in order to quantify consistencies across people in memory performance. At the minimum, there should be at least 40 participants making a memory rating on any given item, although around 80 participants has been shown to result in the most stable memorability scores (Isola et al., 2013). That being said, smaller, targeted studies could be useful (or necessary) when examining memorability patterns in special populations, such as those with Alzheimer's Disease (Bainbridge et al., 2019b).

Another important consideration is whether your stimulus set matches your participant pool—e.g., if you are testing participants within the United States for face memorability, are the face stimuli representative of the diverse demographics of the US (Bainbridge et al., 2013)? This question not only applies for faces, but also for other stimulus types—does it make sense to test participants in Iceland for memory of scene images representing American urban cityscapes?; Does it make sense to test memorability for words in a non-native language? One important note is that all of the research presented in this chapter is tested with participants, laboratories, and stimuli based in the United States. It is still a fascinating and open question how memorability effects may generalize in separate cultural contexts (e.g., is face memorability consistent across countries?). Finally, researchers may want to determine a recruitment sample size with room for error, as they may wish to exclude participants who fail on too many vigilance trials, or who make too many false alarms—both indicators of low attention or random button pressing.

---

[1] The Memorability Experiment Maker: http://wilmabainbridge.com/makeexperiments.html (Bainbridge, 2017).

[2] Amazon Mechanical Turk: https://www.mturk.com/

3. **Memorability scores can be calculated for each stimulus based on memory performance.**

Any measure of memory performance can be used as a "memorability score". Both hit rates (HR; proportion of participants who successfully recognized an image repeat) and false alarm rates (FA; proportion of participants who falsely remembered an item on its first presentation) have been shown to be consistent across observers (Isola et al., 2011; Bainbridge et al., 2013). Different factors can contribute to high HRs or high FAs; for example, an item with high HRs could be one that is highly memorable, or one that tends to cause a lot of responses (both high HRs and high FAs). Images can thus be conceptualized as falling into one of four categories:

1) **High HR, High FA:** "trigger" happy images that evoke a lot of responses, of both accurate memory as well as false alarms.

2) **High HR, Low FA:** memorable images—those where people can accurately recognize an image and have few false alarms.

3) **Low HR, Low FA:** forgettable images—where people do not make false memories, and also do not remember seeing them when they have.

4) **Low HR, High FA:** false memory images—people do not have accurate memories for these images, but somehow make many false memories for them.

All four categories have been shown to be highly consistent; if an image is "trigger happy" for one set of participants, it will likely be "trigger happy" for another set (Bainbridge et al., 2012). While some questions may want to target the study of HR or FA specifically, the two measures can also be combined, with measures such as $d'$ (d-prime, measured by Z-scored HR minus Z-scored FA) and corrected recognition (CR = HR − FA). These measures also show high consistencies across participants (Bainbridge et al., 2013; Bainbridge and Rissman, 2018). Thus, we as experimenters are relatively flexible in what value we decide to take as the "memorability score" for a given image—HR, FA, $d'$, or CR.

4. **Significant memorability score consistency should be validated.**

To ensure memorability scores are meaningful for this set of stimuli (and because it is always important to replicate research findings), researchers should conduct a *consistency analysis* on their set of results (Isola et al., 2011a). Generally, this has been performed by randomly splitting the participant pool into two halves and re-calculating memorability scores within each participant half for each stimulus. Then, a Spearman's rank correlation is calculated between the two halves, to answer the question: Do these two groups of participants remember and forget the same items as each other? These split-halves are conducted across a large number of iterations (usually 1,000-10,000) and the mean Spearman rank correlation is taken as the consistency score. This is compared to a permuted chance level across many iterations, where two participant split-halves are correlated after shuffling their image order. If memory scores are consistent across participants, the split-half consistency should be significantly higher than this permuted null distribution of consistency.

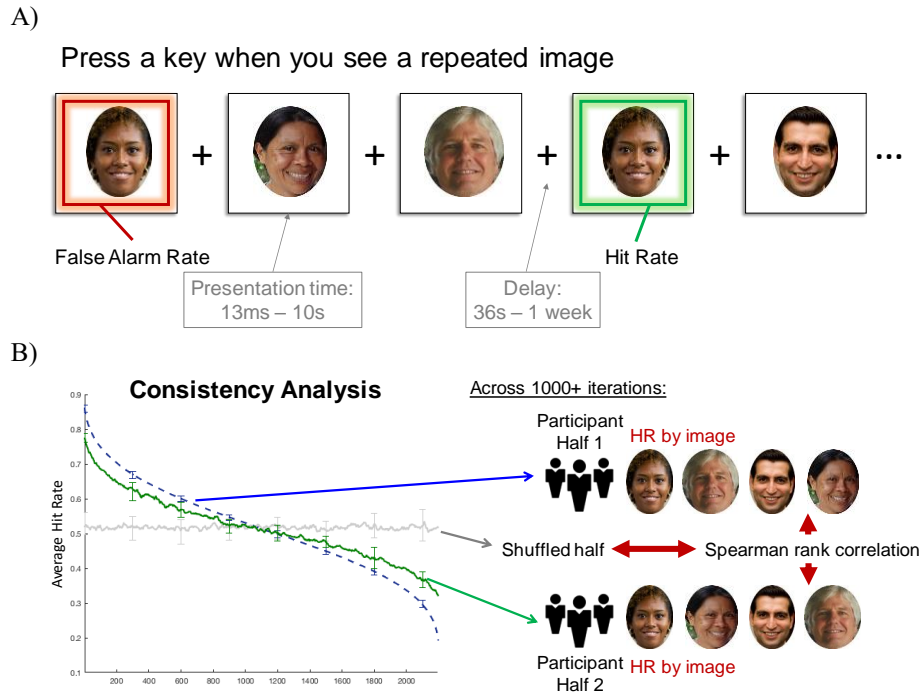

**Fig. 2. The key methods for conducting a memorability experiment.** A) The general methods for a memorability experiment: Participants see a stream of images and press a key when they see a repeat. From these key presses, each image receives a false alarm rate and a hit rate, which both serve as memorability scores. The experimenter has flexibility in their choices in terms of image type, presentation time, and delay between image repeats. B) The split-half consistency analysis to determine how consistent memory performance is across people. Across 1,000+ iterations, participants are split into two random halves and their memorability scores are correlated (Spearman's rank correlation) and compared to a shuffled image order. The graph plots ranked memorability score for participant half 1 (blue dotted line), participant half 2 (solid green line), and the shuffled distribution (gray line).

After these steps, these consistent human-based memorability scores can then be used to answer a myriad of research questions (some of which we outline below).

There have also been in-roads in the realms of computer vision and machine learning to provide automatic quantification of the memorability of an image (Khosla et al., 2013; Khosla et al., 2015) or video (Shekhar et al., 2017; Cohendet et al., 2018). However, using computer-estimated memorability scores in place of ground-truth human ratings should be applied with caution. Some work suggests that current deep learning networks for memorability prediction cannot successfully gener-alize to image sets with more fine-grained category structure or other types of images like faces and visualizations (Squalli-Houssaini et al., 2018), suggesting that these networks may be sensitive to a specific subset of the factors that drive the memorability of an image. For example, deep neural networks trained to predict the memorability for a diverse set of photographs could instead be predicting visual categories of the image that may correlate with memorability but not predict memorability when controlled for (e.g., toy stores may be more memorable than mountain scenes, but can it still find the most memorable mountain?). These algorithms also have been shown to have limited ability to predict memory performance for special populations, such as those with early stages of Alzheimer's Disease (Bainbridge et al., 2019b). However, these limitations could also relate to the training sets of these neural networks; broader and more representative stimulus training sets and more flexible architectures could bridge this gap between human and computer performance (Needell and Bainbridge, 2021).

Instead of using computer-predicted memorability as a replacement for human-measured memorability, the two methods can complement each other in their uses. When studying the psychological and neuroscientific mechanisms that underlie memorability in the human brain, it is essential that ground-truth human scores serve as the main sources of exploration. As it is still unclear what precise information computational models are leveraging to make their predictions, we do not want to create mechanistic claims that could be driven merely by low-level visual or categorical differences. However, computational memorability may also be incredibly helpful for stimulus selection—one could quickly measure the predicted memorability of a stimulus sets to ensure they are roughly controlled for memorability. One could also use these networks to generate images meant to drive strong memory behavior (Goetschalckx et al., 2019). Finally, computational networks assessed against human memory performance could guide development of systems intended to model the human perception and memory systems (Cichy et al., 2019).

## 2.2 Why should we consider memorability?

Now that we have collected memorability scores for a stimulus, what can we do with them? Most straightforwardly, memorability scores are incredibly powerful because they allow for informed predictions about what people will remember or forget. Because stimulus memorability accounts for as much variance in memory performance as all other factors (Bainbridge et al., 2013), this means that incredibly memorable or forgettable stimuli can be selected to drive memory, regardless of the state of the observer. I have shown that memorable images are remembered better than forgettable images no matter how deeply you are attending to and engaging in the images (Bainbridge, 2020). Even if you are just judging whether a fixation cross (+) is black or white, you will remember the irrelevant face behind it if it is memorable. Similarly, even if you are performing a task as deep as judging the honesty of a face, you will still forget it if it is forgettable. I have also shown that reward does nothing to flip these effects—even if you are incentivized with a monetary reward to remember a forgettable image or forget a memorable image, you cannot do either (Bainbridge, 2020). You will still remember the memorable image you were paid to forget over the forgettable image you were paid to remember. We have also found that these memorability scores translate seamlessly across tasks—even if tested with other surrounding image contexts (Bylinskii et al., 2015; Bainbridge et al., 2017), in tasks with a delay (Goetschalckx et al., 2017), or with tasks that only surprise you with a memory test after you study the images (Bainbridge et al., 2017; Goetschalckx et al., 2019; Bainbridge, 2020), the same images emerge as memorable and forgettable. This means that one can design a task with intentionality—to strongly drive memory, have images at both extremes (both memorable and forgettable), or approximate the natural spread of memorability (a Gaussian distribution).

The striking consistency and resilience of this effect has important implications both for the real world and the scientific. In the real world, we can actively design our lives to be more memorable or quantify the memorability of things around us. In a positive sense, we can design educational material to be more memorable to students, or we can design museums to leave a lasting memory on the visitor. In fact, we have found that certain Impressionist paintings at the Art Institute of Chicago are more memorable than others, regardless of their fame or aesthetic value (Hu and Bainbridge, in progress), and such findings could be used to guide the design of exhibit flow, or the creation of new memorable art pieces. We have similarly found some dance movements within the diverse genres of ballet and Korean pop are more memorable than others, and one could envision designing a particularly memorable hit music video, or a memorable sequence for a competition (Ongchoco et al., submitted). It is also important to be aware of potentially negative ways in which memorability could be utilized in the real world—for example, one could make advertisements, slogans, or characters that are unforgettable; and indeed, that is often the original aim of many marketing campaigns (e.g., to make an advertising jingle you cannot get out of your head). It is important that researchers be cautious in how they apply these new principles. We personally vet downloads of our own data to limit access to educational and non-profit research purposes.

Beyond these real-world applications, memorability has important uses for the scientific community. As outlined earlier, it is incredibly easy to get memorability scores for any given set of stimuli. In fact, we have previously used these steps to collect the memorability scores from a decade-old functional magnetic resonance imaging (fMRI) face memory experiment (Rissman et al., 2010), to uncover new decodable patterns of memorability from those data. These results also highlight the fact that memorability exists everywhere, and thus could influence prior findings that did not control for memorability. In the most extreme case, experiments that contrast memory for different stimulus images may be revealing effects entirely driven by the memorability of the images and not the manipulation at hand. For a (yet untested) example, studies contrasting memory performance or brain patterns for familiar celebrity faces versus novel non-celebrities may be eliciting patterns driven by the memorability of the images rather than familiarity—celebrities may generally tend to have more memorable faces (that drive them to become more famous), and so memorable celebrity faces may trigger a separate set of processes from the mixed bag of memorability for the faces of ordinary people. To avoid such concerns about memorability confounding with manipulations of interest, it would be prudent to test the memorability of stimuli in advance when designing an experiment. Many databases with thousands of quantified memorability scores already exist for faces (*10k US Adult Faces Database*: Bainbridge et al., 2013), scene images (Isola et al., 2011; *FIGRIM Dataset*: Bylinskii et al., 2013), objects (*MemCat Dataset*: Goetschalckx et al., 2019b), and abstract visualizations (*MASSVIS Dataset:* Borkin et al., 2013), from which new experimental stimuli could be selected. Taking memorability into account also gives the experimenter power in what sorts of effects they want to see in their study. If they want to eliminate any concerns about memorability acting as a confound with effects of interest, experiments could select stimuli of all medium memorability, or normally or uniformly distributed across the range of memorability scores. Alternatively, if an experimenter wants to drive particularly strong memory effects, they could intentionally select the most memorable, forgettable, and/or false-alarm-able stimuli for their experiments. This can be used to create high-powered experiments, in contrast with traditional memory experiments where it can be difficult to have enough trials that are forgotten or falsely remembered.

Reanalyzing prior data with memorability in mind could also provide fundamental insights into perception and memory that echo beyond this idea of stimulus properties. These insights can come at a very reasonable price, as they do not require the design and collection of new in-lab experimental data. Instead, the stimuli used in a prior experiment on perception, memory, attention, decision making, emotion, social psychology (et cetera) can be put online in a rapid memorability experiment (using the steps outlined previously), and data can be reanalyzed with memorability as an added factor. For example, we have analyzed image-level performance on a previously collected memory test for 394 participants with different levels of dementia (ranging from healthy elderly controls, to cognitive impairments, to Alzheimer's Disease), and found that certain images are more diagnostic than others (Bainbridge et al., 2019b; Bai et al., 2021). Specifically, images that were highly memorable to healthy controls but highly forgettable to those developing dementia were able to better diagnose a held-out set of participants than other image sets of the same size. With these results, one could envision designing highly efficient and brief diagnostic tests for dementia that only use the most diagnostic images. In a therapeutic sense, one could imagine using tools to reinforce memory or provide assistance for particularly forgettable items, or create environments that are intrinsically memorable. As the wealth of open access and Big Data increases with new initiates in the field and resources like the Open Science Framework[3] and the Human Connectome Project[4], it becomes easier to unearth memorability effects from a wide range of experiments. Looking at a diversity of participants, tasks, stimuli, and analytic methods promises to reveal key insights into the underlying mechanisms for memory, and provide a large pool of low-hanging, meaningful, open questions that can be answered without collecting new data.

In fact, research on memorability has already delivered new insights into the human perceptual and mnemonic systems. In exploring the underlying causes of the consistent memorability across people, we can learn more about the factors that drive memory, the processes at the intersection of perception and memory, as well as the mechanisms underlying how the brain stores information in memory. Section 3 discusses our current understanding of what memorability means, and how it relates to a larger psychological framework of human cognition.

## 3. What does it mean for memorability to be an intrinsic image property?

We have demonstrated that memorability is highly consistent across observers, and thus can be conceptualized as a specific property for an image that reliably translates across tasks and observers. This property serves as a measure of memory likelihood—we can use this measure to make predictions about what others will remember or forget. However, it is still unclear what underlying features define an image's memorability, and how it relates to other properties of an image. Is memorability a proxy for another singular image property that is easily image computable (e.g., brightness)? Or, is it some sort of linear combination of several properties of an image? Or, might it represent some more complex interaction of the characteristics of an image? Sections 3.1 - 3.3 explore these three different possibilities for what memorability may represent and how it may be calculated from an image.

### 3.1 Memorability as a singular attribute

When we quantify images, we can think of their properties roughly falling into two different groups: low-level features and high-level features. Low-level features encompass characteristics that can be directly computed from an image without semantic or experiential knowledge, for example, color, brightness, contrast, spatial frequency, and other edge information (e.g., Bainbridge and Oliva, 2015). Conversely, high-level features relate to semantic information such as image category (e.g., "beach") and categorical descriptors (e.g., "natural scene", "outdoor scene"). High-level features also encompass subjective ratings of an image, like aesthetics or emotionality. For the purposes of this chapter, "mid-level" computer vision features (e.g., individual objects) can be conceptually grouped with these high-level features. One early question in the study of memorability was whether it was a low-level or high-level attribute of an image, and whether it merely reflected another already-known stimulus property.

---

[3] The Open Science Framework: http://osf.io
[4] The Human Connectome Project: http://www.humanconnectomeproject.org/

Current evidence shows that memorability is not synonymous with other low-level properties. Image color is not highly predictive of image memorability (Isola et al., 2011), nor is spatial frequency (Bainbridge et al., 2017). Other features that can be extracted from an image, such as its visual saliency, or its gist, have also showed limited explanatory power for memorability (Isola et al., 2011). Additionally, while faces are often quantified by measuring the distances across features, faces with higher differences from the average face are not necessarily more memorable (Bainbridge, 2019). In fact, one can create two sets of images where their average image is indistinguishable, but their images exist at opposite ends of the memorability spectrum (Bainbridge et al., 2017; Bainbridge, 2019).

Memorability also does not serve as a proxy for an alternate high-level property, such as aesthetics. In fact, in spite of an intuition that we may be motivated to remember images we find beautiful, memorability shows relatively low (and negative) correlations with image aesthetics as well as ratings of interestingness (Isola et al., 2013). Similarly, for dance movements, more aesthetic, emotional, or difficult movements are not necessarily more memorable (Ongchoco et al., submitted). In the realm of faces, several face attributes show a correlation with memorability: trustworthiness, kindness, emotionality, atypicality, unfamiliarity, and others (Bainbridge et al., 2013). However, no singular attribute fully captures memorability, and one can have a highly memorable face without having any of these other features. Further, while memorability is highly consistent to a face identity even across viewpoint and expression changes (in other words, a *person* has an intrinsic memorability, not just a singular image of their face), these other attributes do not show significant consistency within an identity (Bainbridge, 2017). Perhaps even more surprising is that for images in general, observers are very poor at predicting what they will remember and forget; there is a non-significant negative correlation between what people think they will remember, and what they actually do remember (Isola et al., 2013). These results highlight the elusive nature of memorability, and how many of our intuitions of what causes these consistencies across people may in fact be false. These results also show that memorability is not reducible down to a single low- or high-level image attribute as far as we know.

## 3.2 Memorability as a combination of attributes

It thus seems clear that memorability is not capturing a low-level or high-level property already used to quantify images. However, perhaps a combination of these properties can be used to successfully predict memorability, just as other high-level attributes of an image can be broken down into a combination of several properties. For example, although it is often considered and rated as a single dimension, facial attractiveness can be quantified as a combination of symmetry, skin quality, youthfulness, and cultural templates of attractiveness (e.g., Perrett et al., 1999; Schmid et al., 2008). Is memorability similarly distillable into a series of properties that can be used to quantify an image? In the realm of faces, a LASSO regularized regression including twenty attributes to quantify faces (e.g., attractiveness, trustworthiness, dominance, intelligence) and memories (e.g., typicality, commonality, subjective ratings of memorability) was only able to explain less than half of the variance in memorability scores (Bainbridge et al., 2013). For more complex stimuli like dance movements, a set of ten attributes including beauty, emotionality, complexity of movement, speed, difficulty, atypicality, and subjective memorability only captured 6.40% of the variance in memorability scores (Ongchoco et al., submitted). And, for more semantically-based stimuli like words, attributes commonly used in linguistics like the concreteness or frequency of a word did not explain unique variance for the memorability of a word (Xie et al., 2020). These combinations of attributes are able to better predict memorability than a singular attribute alone, and the relative weighting on each property can reveal what types of information have more impact on what people ultimately remember. However, the fact that none of these combined models can explain much of the variance in memorability suggests that memorability exists as something beyond a linear combination of various features.

## 3.3 Memorability as an arrangement of attributes

Rather than representing a linear combination of specific properties, memorability could instead represent an arrangement across properties. For example, rather than it being that more attractive faces are more memorable, instead it could be that more *extreme* faces (highly attractive *and* highly unattractive) are more memorable. Images can be conceptualized along a multi-dimensional space, where each dimension represents an attribute and a single image can be represented as a point, located by the vector made up of its score on each of these attribute dimensions. A highly atypical or distinctive image with extreme values on its attributes would be on the outskirts of this distribution, while a more typical or common image with attributes near the mean or prototype would exist near the center of this distribution. The clustering structure of items in a set could also be predictive of memorability, where items with representations similar to others may be more forgettable, where items with more distinctive representations may be more memorable. Prior work has suggested that atypical or distinctive faces tend to be the most memorable (Light et al., 1979; Winograd, 1981; Bartlett et al., 194; Vokey & Read, 1992), and perhaps these same intuitions apply more generally across memorability for images.

Indeed, there is some converging evidence that memorability may represent the location of a stimulus in such an attribute-based distribution. In the realm of images, images that are located in sparser areas of the attribute space (as defined by features extracted from convolutional neural networks) tend to be more memorable (Lukavský and Děchtěrenko, 2017). However, it is still an open question what attributes constitute such a space, whether they contribute in equal weight, and whether there are separate influences from low-level visual features, versus high-level semantic information. Some work specifically looking at scene images proposes that high-level similarities (i.e., being of the same scene category) but low-level dissimilarities (i.e., having large color differences) may be most correlated with memorability (Koch et al., 2020). However, memorability effects still occur for stimulus categories that are relatively matched for low-level visual information and contain the same level of semantic information, like faces (Bainbridge et al., 2013). Conversely, memorability effects also still occur for semantically rich and diverse stimuli with relatively low visual information, like concrete nouns (Xie et al., 2020). So it is still unclear if these principles of similarity drive memorability similarly across stimulus types.

Perhaps the most convincing evidence for memorability as a reflection of a representational space comes from the realm of words and computational linguistics. The Global Vectors for Word Representation (GLoVE) model (Pennington et al., 2014) is a model that creates a vector for a given word representing its relationship to other words. GLoVe characterizes words by their co-occurrences in real-world text, with the idea that words that tend to co-occur in a sentence tend to be more semantically related. For example, the word "foot" may frequently co-occur in a sentence with many words, like "shoe", "hand", or "ball"; in contrast, a word like "dime" may co-occur with fewer words (e.g., "coin"), and thus have fewer semantic connections as well. If this GLoVE model for semantic relatedness captures how we represent words in memory, then the network structure could show some relationship to memorability. For example, it could be that these highly interconnected words (or "roots" of the network) are the most memorable items, while the most sparsely connected words (or "leaves") are the most forgettable items. In fact, this is what we observe if we make predictions of memorability based on this network structure (Xie et al., 2020); in our dataset of 300 concrete nouns, "foot" was the most memorable (96.3% of online observers successfully recalled it), while "dime" was the most forgettable (only 16.2% successfully recalled it). In a multiple regression model, the GLoVE score for a given word explained significant, unique variance in memorability for that word, while other linguistic attributes such as concreteness (Brysbaert et al., 2014) and word frequency (Davies and Gardner, 2013) explained no additional unique variance. Memory behavior also matched many of the predictions one would make if the roots of the network are more memorable than the leaves. First, one would predict that during the retrieval process, you would visit the roots first and then traverse the network to reach the outer leaves; indeed, we observe that retrieval speed is faster for memorable items than forgettable ones. When failing to retrieve an item, one would also predict that intrusions (falsely recalled items) would most commonly be these roots first visited in the retrieval process. Indeed, intrusions had significantly higher memorability than the median memorability of the set.

The combination of these findings suggests that memorability reflects the position of an item in our mental network of knowledge and prior experiences. Rather than being driven by any one attribute or combinations of attributes, it is instead an item's location in a larger network of items arranged by similarity across these attributes that determines an item's memorability. I will discuss what this might mean about the brain in Section 4.

However, there is some initial evidence that the relationship between the stimulus space and memorability may not be uniform across all stimuli (Kramer et al., 2021). For example, initial evidence suggests that within some object categories, highly atypical items that are dissimilar from other within-category items are memorable (e.g., an atypical kitchen appliance, like a hearth, is more memorable than a stove), while for other object categories, highly typical items that are similar to other within-category items are memorable (e.g., a typical weapon, like a pistol, is more memorable than a tank). In fact, while studies on images have suggested that visually distinctive and sparsely distributed items are more memorable (Lukavský and Děchtěrenko, 2017; Koch et al., 2020), studies on words have suggested that semantically connected items are more memorable (Xie et al., 2020). There is also evidence that the predictions of memorability for real-world scene images derived from a deep learning neural network correlate with memory patterns in the rhesus macaque (Jaegle et al., 2019), even though these monkeys have not experienced a wide range of scenes that would allow them to construct these mental networks of similarity. Thus, while the idea that memorability reflects a network-like arrangement of mental representations is an attractive one, there are still many open questions about what exact measure memorability reflects, and how it can be predicted regardless of stimulus category (faces, scenes, words) or granularity of that category (scenes versus kitchens).

With this working hypothesis that memorability is defined by an image's relationship to other images, we can now delve into the brain and explore whether we find supporting evidence. Moreover, we can look under the hood to see how processing of memorability may relate to other neural processes, like vision, attention, and memory encoding.

# 4. The brain mechanisms underlying memorability

Thus far, we have discussed memorability as a property of a stimulus—stimuli are intrinsically memorable or forgettable, and this influences the likelihood for an observer to remember that image. However, memorability is also tied to a pair of cognitive processes triggered by the image: the process of perception, and the process of memory encoding. Thus, when we see a memorable (or forgettable) item, what patterns may we observe in the brain?

There are four main hypotheses that would predict how memorability should influence processing in the brain. First and most uninterestingly, we may see nothing in the brain when looking for a "signature of memorability" (neural differences between memorable and forgettable images; Section 4.1). Memorability appears to be a complex property, irreducible to a simple set of other measures. Many high-level image properties do show patterns in the brain, such as whether a scene image is natural or manmade or open or closed (Kravitz et al., 2011; Park et al., 2011), and whether the objects in it are big or small (Konkle and Oliva, 2012). However, not all properties we respond to behaviorally show a strong signal in the brain, as the ability to detect a signal usually requires high, consistent signal across a localized set of neurons, replicable across participants. These signals must also be prolonged enough and in cortical areas accessible enough to be detectable by a method like fMRI, which has low temporal resolution and high but imperfect spatial resolution (one unit of the brain—a voxel—still can reflect the average of the firing of 100,000 neurons). However, if we do observe differences between memorable and forgettable images in the brain, a second hypothesis would be that that memorability effects could be reflected in visual processing (Section 4.2). Since memorability is an intrinsic property to a specific image, regardless of the observer, it seems that it could reflect a perceptual calculation that would be reflected during early vision, or heightened attention to a visual feature (Section 4.3). A third hypothesis is that we would see memorability effects during memory encoding (Section 4.4). As memorability scores are operationalized by successful memory encoding and retrieval, we may see effects much like those observed during other memory encoding tasks at the individual subject level. Finally, a fourth hypothesis is that we uncover a new pattern in the brain related to memorability, as some sort of intermediate step between perception and memory (Section 4.5).

## 4.1 Can we find memorability in the brain?

The first investigation of memorability in the brain tested whether there were any differences between viewing memorable images and forgettable images in the first place (Bainbridge et al., 2017). Sixteen participants in an MRI scanner viewed a stream of images and performed a relatively simple but fast-paced perceptual task— when they saw a face, they had to quickly categorize its gender (male/female), and when they saw a scene, they had to quickly categorize its location (indoor/outdoor). Unbeknownst to these participants, half of the faces and scenes were highly memorable, while the other half were highly forgettable—determined in advance from online memorability studies (Isola et al., 2011; Bainbridge et al., 2013). Furthermore, these stimuli were controlled for a range of potential attributes that could be confounded with memorability. The memorable and forgettable faces were matched on low-level properties like spatial frequency and color, high-level properties like ratings of their emotionality, attractiveness, friendliness, and confidence, and demographics like their gender, race, and age. The memorable and forgettable scenes were matched on low-level properties like spatial frequency, color, number of objects, and size of objects, and on high-level properties like whether they were indoor or outdoor, natural or manmade, and had no people or animals in them. Thus only the memorability of these items differed between conditions. Participants also were not aware of any memory-related nature of the task and so would not utilize any honed mnemonic strategies. Instead, they were given a surprise memory test only after getting outside the scanner.

We discovered that when contrasting memorable versus forgettable images, there was in fact a significant, strong, bilateral swath of heightened signal for memorable images, bridging from late visual areas to subcortical mnemonic areas (Figure 3). It appeared that memorability effects do exist in the brain. And, these effects were not driven by other properties (because we controlled for all other properties known to correlate with memory), nor by intentional mnemonic strategies (as participants believed this was a purely visual task). The next question was how these effects related to those of other cognitive processes.

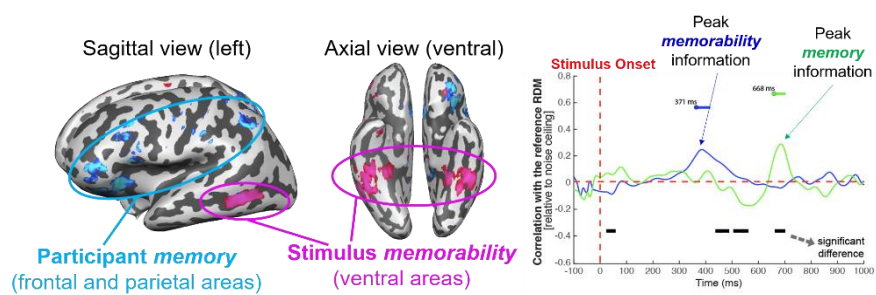

**Fig. 3. Neuroscience findings of memorability.** Utilizing fMRI (left and center), we have replicated across multiple experiments a bilateral swath of memorability patterns along the ventral visual stream extending into the medial temporal lobe (Bainbridge et al., 2017; Bainbridge and Rissman, 2018). In contrast, patterns related to an individual's memory have been identified in frontal and parietal areas. Utilizing MEG (right), we have also identified a temporal separation of these processes, where memorability-based information is discriminable after early vision, but before memory encoding processes (Khaligh-Razavi et al., 2016).

## 4.2 Memorability and the visual system

In the brain during perception, incoming visual information generally traverses a path (the "ventral visual stream") that begins at areas for processing low-level visual information like edge eccentricity and orientation (i.e., early visual cortex, EVC), and then high-level information like stimulus category is parsed in downstream regions along the inferotemporal cortex (Kravitz et al., 2013). Just as one key psychological question is how memorability relates to low-level and high-level properties for an image, a similar neuroscientific question is how this sensitivity to memorability in the brain relates to early and late visual regions.

In our studies of memorability, we have consistently failed to observe effects of memorability in early visual cortex (Bainbridge et al., 2017; Bainbridge and Rissman, 2018), suggesting that these effects are not driven by low-level visual features. In contrast, several pieces of evidence seem to suggest that memorability may relate to higher order visual processes. Memorability effects for images have been repeatedly observed in late visual areas along the inferotemporal cortex, both during memory encoding (Bainbridge et al., 2017) as well as during retrieval (Bainbridge & Rissman, 2018). For a cued word recall task examining memorability in the absence of visual input, memorability signals were instead present in semantic areas such as the anterior temporal lobe, rather than perceptual areas (Xie et al., 2020). In the temporal domain, the timing of memorability decoding occurs around the same time as other late perceptual processes, at around 150-400ms after stimulus onset (Khaligh-Razavi et al., 2016; Mohsenzadeh et al., 2019), and before early visual processes at around 100ms. These results suggest that the brain shows sensitivity to the memorability of a stimulus during late perceptual processes.

## 4.3 Memorability and attention

However, some counterintuitive findings have also emerged from these late visual areas. For example, the fusiform face area (FFA) is a localized region that shows high sensitivity to face images but not scenes, while the parahippocampal place area (PPA) conversely shows high sensitivity to scenes but not faces. Bainbridge et al. (2017) observed that both regions showed sensitivity to the memorability for both the preferred and un-preferred category; for example, one could detect scene memorability from the FFA. These results could suggest that increased brain activation related to stimulus memorability is partially due to attention-driven boosts in signal. For example, heightened attention to memorable stimuli may result in increased activation across visual areas in comparison to forgettable stimuli, resulting in these effects even across category-selective areas.

While the interaction of attentional state and memorability is still an open question, various evidence suggests that memorability effects cannot be solely explained by heightened attention. A series of behavioral studies found that memorable items do not necessarily capture attention (i.e., bottom-up attention); your eyes (and your spotlight of attention) are not automatically drawn to memorable items when searching for a target (Bainbridge, 2020). Similarly, intentional efforts to control attention (i.e., top-down attention) also do not affect memorability; even with highly attentive tasks or rewards to drive your attention, you will always remember memorable items better than forgettable ones (Bainbridge, 2020). In recent work directly contrasting attentional state and memorability, we have observed no correlation between both factors, and they contribute unique variance to predictions of ultimate memory behavior (Wakeland-Hart et al., 2020). Differences in activity for memorable versus forgettable images has also not been identified in areas typically associated with attention, such as regions within the frontal and parietal cortices (Culham et al., 2001). However, future work will need to investigate further the links between attention and memorability during the memory encoding process.

## 4.4 Memorability and the memory system

Patterns for memory in the brain have been frequently defined using a method looking at "subsequent memory", in which trials during encoding are sorted post-hoc based on whether they were subsequently remembered or not (Brewer et al., 1998; Wagner et al., 1998). This contrast theoretically reveals which brain regions show activity based on an item's memory fate (i.e., whether it will be successfully encoded and maintained in memory), and has often identified areas in the frontal and parietal lobes (Rissman et al., 2010; Kim, 2011). Memorability could potentially serve as another way to access this subsequent memory signal—since memorable items tend to be subsequently remembered, perhaps the contrast of memorable vs. forgettable will be identical to that for remembered vs. forgotten. However, surprisingly, we find a dissociation between these two effects; regardless of whether an item is remembered or forgotten by a participant, their brain still shows sensitivity to its memorability (Bainbridge et al., 2017). And, patterns for these two contrasts of information suggest distinct cortical areas, with memorability sensitivity in ventral areas, and subsequent memory in frontal and parietal areas (Bainbridge and Rissman, 2018). These results suggest distinct processes related to *stimulus* memorability versus *subject* memory. Indeed, temporally resolved methods show that memory encoding (identified by differentiable signal for subsequently remembered versus forgotten items) occurs after patterns of memorability; early vision occurs early around 100ms, then sensitivity to memorability at 150-400ms, and then memory encoding at 600-800ms (Khaligh-Razavi et al., 2016). Thus, this sensitivity to memorability in the brain is not due to the process of memory encoding, but some sort of signal that occurs *before* encoding.

## *4.5 Memorability as a prioritization signal*

It thus seems clear that the brain is sensitive to the memorability of an item, and not because memorable items are visually distinct, nor because they elicit heightened attention, nor because of a difference in successful encoding. Instead, this sensitivity to memorability comes online during an intermediate time point between early vision and memory encoding (Khaligh-Razavi et al, 2016). Sensitivity to memorability has also been identified in memory-related areas in the medial temporal lobe, like the hippocampus and perirhinal cortex (PRC). The PRC has been considered by some researchers to act as a novelty detector (Desimone, 1996; Brown & Aggleton, 2001; Daselaar et al., 2006), but perhaps it may be sensitive to subtler statistical differences across items, like their memorability. In the word domain, temporal lobe structures (specifically, the anterior temporal lobe) also show a sensitivity to memorability in the absence of a perceptual experience of that item (Xie et al., 2020). Thus, memory-related temporal areas may be the seat where this memorability signal originates, even if the signal differs from those of memory encoding.

Why would memorability be "computed" by any particular region of the brain? Our current hypothesis is that memorability represents a *priority signal* for a stimulus. We cannot remember everything we perceive—our memories are limited from moment to moment. Thus, our brains must rapidly and efficiently sort information for memory encoding. Perhaps these findings on memorability have detected this sorting signal, where high priority items are those with certain statistical characteristics that recommend they be encoded into memory, while low priority items are safe to be forgotten. This interpretation is highly suggested by the connection between semantic networks and word memorability, where high-priority and highly memorable items are those items where we start our memory searches, with the highest semantic connections (Xie et al., 2020). However, this pattern is not only found in the domain of words; for images as well, we see a distinctive concentric pattern of memorability in late visual and memory areas. Specifically, memorable images are neurally similar to each other, while forgettable items are more dispersed. These patterns suggest that memorable items do not just trigger higher signal in the brain, but a patterned signal, with memorable items at the roots of a larger structure (Figure 4).

This hypothesis is still new but largely testable. We can design tasks to manipulate the priority of an item and see how this influences memory, by examining how patterns of memorability emerge for novel stimulus categories with designed statistical relationships. We can also see how memorability relates to other concepts like processing fluency or information density. These findings promise to guide meaningful exploration in the domains of computer science as well; can memorability of an item be used as a guiding principle for designing user interfaces, or human-computer interactions, whereby computer systems sort information or behaviors by their need to be remembered?

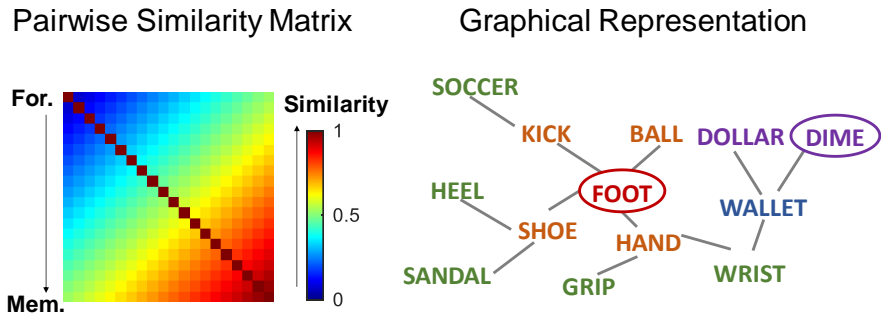

**Fig. 4. Depiction of memorability as a priority signal.** The left depicts the pairwise representational similarity matrix that has been repeatedly observed in the brain (the areas shown in Fig. 3). If the correlations of brain patterns are calculated between each pair of stimuli, then memorable items show high similarity, while forgettable items show low similarity to each other and other items. This suggests a mental representation like the one on the right, where memorable items are more centralized and forgettable items are more dispersed. This shape is also suggested by the GloVe model and the findings of Xie et al. (2020), where more memorable items (e.g., "foot") have a larger number of semantic connections and are located at the roots of a network, in contrast with more forgettable items (e.g., "dime") which have fewer connections and are located at the leaves of the network.

## 5. Conclusion

The study of image memorability is a nascent but promising topic, with important implications for the fields of human perception and human memory, as well as implications for computational models of these processes. In Section 2, I motivated the necessity to consider memorability in both the scientific and the real worlds, and presented practical tips and tools so the reader can dive into their own studies of memorability (even just using their own pre-existing data). In Section 3, I discussed memorability's role as an image property, and its relationship to other low- and high-level image attributes. Finally, in Section 4, I delved into the neural mechanisms underlying memorability effects, and how they could reflect memorability's role as a prioritization signal.

We currently know more about what memorability *isn't* than what it is. We know it is not just a stand-in for another already known image attribute, or a combination of other pre-existing attributes. We know memorability signals in the brain are not mere reactions to visual differences, or signals of heightened attention, or patterns of memory encoding. In terms of what we do know, we see evidence that sensitivity to memorability occurs somewhere between perception and memory, and we take this as preliminary evidence that it might reflect how we sort (or prioritize) perceptual information for memory encoding. A key future direction will be the creation of generative models of memorability, whether developed through psychophysics, neuroscience, or computational models. The ability to create honed predictions about stimulus memorability will in turn allow for precise predictions of individuals' memories, which will have resounding impacts in both the psychological and applied realms.

There are many exciting future directions from which we can continue the study of memorability. All of the work I have discussed has been in the visual domain, with a large focus on static images. Future work will need to elaborate the memorability of stimuli across dynamic episodes (e.g., interactions, conversations), modalities (e.g., olfaction, audition), and across items (e.g., across associations and contexts). These findings also highlight the importance of considering the many factors that influence a behavior; while the system performing the behavior (the human) is important to consider, so are the inputs to that system as well (e.g., Bainbridge et al., 2020). Beyond memorability, a deeper look into stimulus properties, their consistencies (or differences) across people, and their related behavioral phenomena may spur many valuable directions of inquiry. One related and equally intriguing stimulus property we are now beginning to investigate is the propensity for an image to trigger false memories—certain images cause the same visual false memories across people even when these false versions have never been visually experienced (dubbed the *Visual Mandela Effect*; Prasad and Bainbridge, 2021). More broadly, these stimulus-centric investigations have high potential for providing new insights into the mechanisms of human cognition, by understanding how specific inputs trigger different behavioral outputs.

In sum, memorability promises to revolutionize our understanding of visual memory. It has started to reveal the intricate processes occurring in the brain between the perception and encoding of an item. It also presents a way to measure and control memories through the items we are remembering. It is with these explorations into the items we see that we may hope to take agency over our memories.